\documentclass[aps,pra,reprint,groupedaddress,amsmath,amssymb,superscriptaddress,twocolumns]{revtex4-1}
\usepackage{graphicx}
\usepackage{amsmath}
\usepackage{dcolumn}
\usepackage{bm}
\usepackage{hyperref}
\usepackage{xcolor}
\usepackage{lipsum}
\usepackage{cancel}
\usepackage{mathtools}
\usepackage{float}
\usepackage{centernot}
\usepackage{soul}
\graphicspath{{./Images/}}
\begin{document}
\title{Strong coupling diagnostics for multi-mode open systems} 
\author{C. Kow}
\affiliation{Department of Physics and Applied Physics, University of Massachusetts, Lowell, MA 01854, USA}
\author{Z. Xiao}
\affiliation{Department of Physics and Applied Physics, University of Massachusetts, Lowell, MA 01854, USA}
\author{A. Metelmann}
\thanks{Present address: Dahlem Center for Complex Quantum Systems and Fachbereich Physik, Freie Universit\"{a}t Berlin, 14195 Berlin, Germany}
\affiliation{Department of Electrical Engineering, Princeton University, Princeton, NJ 08544, USA}
\author{A. Kamal}
\affiliation{Department of Physics and Applied Physics, University of Massachusetts, Lowell, MA 01854, USA}
\date{\today}
\begin{abstract}
We present a new method to diagnose strong coupling in multi-mode open systems. Our method presents a non-trivial extension of exceptional point (EP) analysis employed for such systems; specifically, we show how eigenvectors can not only reproduce all the features predicted by EPs but are also able to identify the physical modes that hybridize in different regions of the strong coupling regime. As a demonstration, we apply this method to study hybridization physics in a three-mode optomechanical system and determine the parameter regime for efficient sideband cooling of the mechanical oscillator in the presence of reservoir correlations.
\end{abstract}
\maketitle
%
\section{Introduction}
%
Strongly-coupled open systems form the operational framework in diverse fields, ranging from quantum information processing, precision measurements to quantum chemistry. One of the main challenges in modeling such systems is the appearance of strongly-hybridized dressed states beyond a critical coupling strength, which necessitates describing dissipative dynamics in a non-local basis. A powerful framework for analyzing this transition from weak to strong coupling in open systems is provided by exceptional points (or EPs). EPs are branch point singularities in the parameter space, where two (or more) eigenvalues and eigenstates of the system coalesce. This makes them distinct from degeneracy points in Hamiltonian systems, which support identical eigenvalues while corresponding eigenvectors remain orthogonal. The physics of EPs continues to be exploited in a variety of applications involving non-Hermitian physics, such as novel nonreciprocal devices \cite{Peng2016,Yoon2018,Hassani2018} and amplifiers \cite{Choi2017,Zhong2020}, quantum sensors \cite{Wiersig2014,Chen2017,Zhang2019}, and single-mode lasers \cite{Feng2014,Hodaei2014,Peng2014} to name a few.
\par
Though EPs represent points where both eigenvalues and eigenvectors collapse to a single value, the analysis and design of open systems utilizing EPs predominantly makes use of eigenvalues of the dynamical matrix \cite{Seyranian2005}. This is rooted in the fact that the non-trivial topological properties associated with the emergence of such degeneracies, such as non-adiabatic mode switching \cite{Milburn2015} and chiral state transfer \cite{Xu2016}, can be entirely described by tracking the eigenvalues alone in the complex parameter space \cite{Heiss1999}. In this paper, our focus is quite different: rather than study the properties of the dressed states, we aim to study the strong-coupling physics from the point-of-view of physical subsystems. To this end, we present a new method that shows how eigenvectors can provide a comprehensive description of strong coupling effects in open systems. The basic idea relies on exploiting the mode correlations as reflected by the eigenvector projections in relevant subspaces of an $N$-dimensional mode space. Our proposed method can not only reproduce all the features obtained from eigenvalues, but provide more nuanced information about different types of correlations in a multi-mode open system under strong coupling. Most importantly, it provides a means to identify the physical modes that hybridize to form the dressed eigenstates (also referred to as `supermodes'), a feature not accessible with eigenvalues. We emphasize the physical significance of such subsystem identification in strong-coupling manifolds, using the example of cooling of a mechanical oscillator to its quantum ground state using engineered dissipation. The proposed criterion enables characterization of the operational cooling regime, where the mechanics remains weakly coupled to a multi-mode reservoir.
\par
The paper is organized as follows: we begin with a description of an $N$-mode open system with nearest-neighbor interactions in Sec.~\ref{Sec:evecs}, and use $N=3$ and $N=4$ cases as examples to illustrate the inadequacy of conventional eigenvalue-based EP analysis when extended to more than two modes. We then introduce the eigenvector projection-based method in Sec.~\ref{Sec:evecs} and show how it can be used to generate the detailed coupling map of a multi-mode open system, resolving the shortcomings of the usual EP analysis. In Sec.~\ref{Sec:cooling}, we examine quantum ground state cooling in a three-mode optomechanical system to show how the proposed method can be applied to a physical problem of interest. We conclude with a summary of main results and offer perspectives for potential extensions of our study in Sec.~\ref{Sec:conclusions}. Additional calculations details are included in appendices \ref{App:Evalanalyis} and \ref{App:full3mode}.
%
%
\section{Exceptional points in a multi-mode system}
\label{Sec:evals}
%
A generic $N$-mode open system with nearest-neighbor hopping interactions can be described by a Hamiltonian of the form,
\begin{eqnarray}
    \mathbb{H}^{(N)} = \sum_{\substack{j=1,k=1\\\langle j, k\rangle}}^{N}\left(\frac{\Delta_{j}}{2}\delta_{j,k} + g_{jk}(1-\delta_{j,k})\right)a_{j}^{\dagger}a_{k},
    \label{Eq:Hamiltonian}
\end{eqnarray}
written in the interaction frame defined with respect to the Hamiltonian $\sum_{j}\omega_{j}^{d}a_{j}^{\dagger}a_{j}$, with $\Delta_{j} = \omega_{j} - \omega_{j}^{d}$ being the detunings associated with each mode. The phase of the couplings is determined by $\arg(g_{jk})$, with $g_{jk} = g_{kj}^{*}$ ensuring hermiticity of the interaction Hamiltonian. The open dynamics of this system can be derived from Heisenberg-Langevin equations for the mode annihilation operators, $a_{j}$, as
\begin{align}
    \frac{d{\bf V^{\text{(N)}}}}{dt} = \mathbb{M}^{\text{(N)}}{\bf V^{\text{(N)}}} + \sqrt{\mathbb{K}^{\text{(N)}}}{\bf V}^{\rm in (N)},
    \label{eq:QLE}
\end{align}
where  ${\bf V}^{\rm (N)} = [a_1, a_{2},...,a_{N}]^{\rm T}$, ${\bf V}^{\rm in(N)} = [a_1^{\rm in}, a_{2}^{\rm in},...,a_{N}^{\rm in}]^{\rm T}$ denote the internal mode and input noise operators respectively, and ${\mathbb{K}^{\text{(N)}}= {\rm diag}(\kappa_{1}, ...\kappa_{j}, ..., \kappa_{N})}$ is a diagonal matrix with its non-zero elements representing the decay rates associated with each individual modes. The dynamical matrix $\mathbb{M}^{\text{(N)}}$, also referred to as the ``mode matrix", for the system with nearest neighbor couplings considered here is an ${N \times N}$ tridiagonal complex-symmetric matrix of the form
\begin{eqnarray}
    \mathbb{M}^{\text{(N)}} = 
    \left(
    \begin{array}{ccccc}
    \widetilde{\Delta}_{1}& - ig_{12} & 0  & \ldots & 0\\
    - i g_{21} & \widetilde{\Delta}_{2} & - ig_{23}  & \ldots  & 0\\
    0 & - ig_{32} & \widetilde{\Delta}_{3} & \ldots & 0\\
    \vdots & \vdots & \vdots & \ddots & \vdots\\
    \end{array}
    \right),
\end{eqnarray}
where $\widetilde{\Delta}_{j} \equiv \Delta_{j} - i \kappa_{j}/2$. Note that here we have assumed open boundary conditions; closed-loop topologies with periodic boundary conditions have been studied in the past and while they can support qualitatively new physics, the shape of coupling map is not germane to the question of diagnosing strong coupling that we focus on in the following sections.
\begin{figure}[t!]
\centering
\includegraphics[width=\columnwidth]{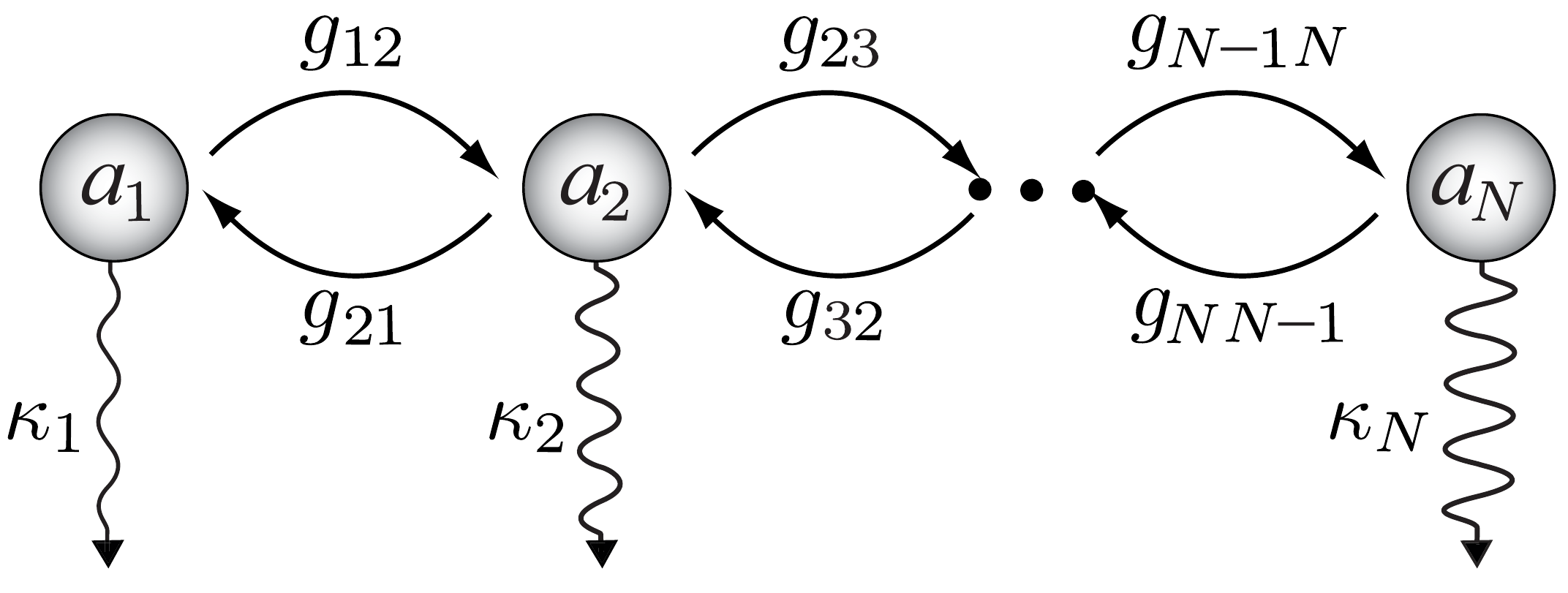}
\vspace{-10pt}
 \caption{Schematic of an $N$-mode open system with nearest neighbor interactions. Curved arrows depict the bilinear interactions $g_{ij}$, while local decay rates of the modes are depicted with $\kappa_j$.}
 \label{Fig:multimodesys}
\end{figure}
\par
\begin{figure*}[t!]
\centering
\includegraphics[width=0.9\textwidth]{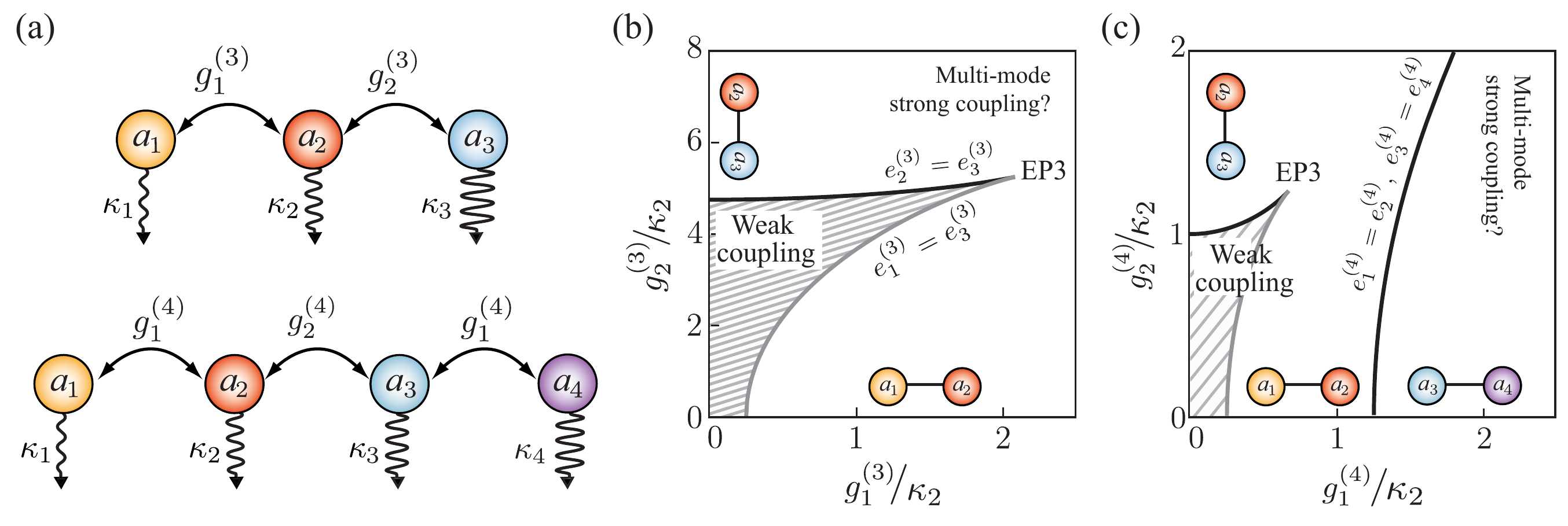}
\vspace{-10pt}
\caption{(a) Illustration of three-mode (${N=3}$) and four-mode (${N=4}$) systems with nearest-neighbor hopping interactions. The hopping interactions are depicted with double-headed arrows $g_j^{(N)}, j \in [1,N-1]$, and local decay rates are represented as wavy arrows $\kappa_j, j\in[1,N]$. (b), (c) Coupling phase diagrams for ${N=3}$ and ${N=4}$ systems obtained using eigenvalues of the respective mode matrix.  The decay rates used in the calculation are $\kappa_1/\kappa_2 = 0.01$ and $\kappa_3/\kappa_2 = 20$ for ${N=3}$ case, and $\kappa_1/\kappa_2 = 0.01$,  $\kappa_3/\kappa_2 = 5$ and $\kappa_4/\kappa_2 = 10$ for ${N=4}$ case. In each plot, the solid curves denote the locus of EP2s in the parameter space, calculated using Eq. (\ref{Eq:EP2disc}), and are labeled with the eigenvalues that coalesce at the respective EP. The hatched (white) region correspond to the weak (strong) coupling regime. The insets near the axis pictorially depict the relevant pair of modes that hybridize when the coupling strength is increased beyond the corresponding EP2 threshold.}
\label{Fig:OpenLoop}
\end{figure*}

Conventionally, weak and strong coupling regimes are identified by finding the exceptional points (EPs) supported by $\mathbb{M}^{(N)}$. For instance, for the well-known case of two-modes coupled with a hopping-type interaction, an EP2 is realized for $g^{(2)}_{\text{EP2}}=|\kappa_1-\kappa_2|/4$. In the weak coupling regime, with $g<g^{(2)}_{\text{EP2}}$, the eigenvalues are purely real while in the strong-coupling regime, with $g>g^{(2)}_{\text{EP2}}$, the eigenvalues become complex; the imaginary part corresponds to the detuning of the mode from resonance due to hybridization that lifts the degeneracy, manifesting as a ``splitting" of the mode spectrum.  In general, this transition between real and complex solutions (or EP2) for an $N$-mode system can be obtained by setting the discriminant of the characteristic polynomial of the mode matrix, $p_{\mathbb{M}^{(N)}} = {\rm det}(\lambda \mathbb{I} - \mathbb{M}^{(N)})$, to zero, 
\begin{equation}
   {\rm disc}(p_{\mathbb{M}^{(N)}}) \equiv \Pi_{\alpha \neq \beta} (\lambda_{\alpha} -\lambda_{\beta}) = 0,
   \label{Eq:EP2disc}
\end{equation}
where $\lambda_{\alpha, \beta}$ denote a pair of eigenvalues \footnote{Technically, $p_{\mathbb{M}^{(N)}}$ needs to be the minimal polynomial of the mode matrix.}. Since $p_{\mathbb{M}^{(N)}}$ is a polynomial of degree $N$ in $\lambda$, the strong coupling regime needs to be studied in a hyperplane spanned by $N-1$ coupling parameters $g_{ij}$ for fixed values of decay rates $\kappa_{j}$.  As concrete examples, we now consider $N = 3$ and $N=4$ systems depicted in Fig.~\ref{Fig:OpenLoop}(a) in detail, and describe the generic features of EPs in systems with bilinear interactions. 
%
%
\subsection{$N=3$ case}
%
Time-averaged dynamics of a three-mode open system with nearest neighbor couplings can be described by a $3\times 3$ mode matrix of the form,
\begin{equation}
\mathbb{M}^{(3)} = 
\left(\begin{array}{ccc}
                     -\kappa_1/2 & -ig_{1}^{(3)} & 0  \\
                    -ig_{1}^{(3)} & -\kappa_{2}/2 & -ig_{2}^{(3)} \\
                     0 & -ig_{2}^{(3)} & -\kappa_{3}/{2}
                \end{array} 
\right) \label{Fig:ModeMat3},
\end{equation}
where $g_{12} = g_{1}^{(3)}, g_{23} = g_{2}^{(3)}$, with $g_{j}^{(\rm{3})}\in\mathbb{R_{>\text{0}}} \, \forall j$. Here, without loss of generality, we have considered resonant driving, leading to zero detunings, i.e. $\Delta_{j} = 0, \, \forall j$. Fig.~\ref{Fig:OpenLoop}(b) shows a plot of EP2s for this system as a function of the interaction strengths, obtained from Eq. (\ref{Eq:EP2disc}) for fixed values of decay rates $\kappa_{j}$. Analogous to the two-mode setup, we note that the EP2s demarcate the weak and strong coupling regimes; specifically, the intercepts on the x and the y axes correspond to $\gamma_{-} = |\kappa_{1}-\kappa_{2}|/4$ and $\kappa_{-}= |\kappa_{2}-\kappa_{3}|/4$ respectively, which are the EP2 thresholds for decoupled $\{a_{1},a_{2}\}$ and $\{a_{2},a_{3}\}$ subsystems in the absence of other couplings. We emphasize that within the region bounded by the EP2 curves, the system is in the weak coupling regime with purely real eigenvalues, whereas outside this region all three eigenvalues can be complex and the system is in the strong-coupling regime. 
\par
A noteworthy feature for systems with $N>2$ is the appearance of higher-order exceptional points. For instance, as shown in Fig.~\ref{Fig:OpenLoop}(b), a three-way exceptional point (EP3) is realized at the coincidence of two EP2 curves, where all three eigenvalues and eigenvectors become identical. The coordinates of EP3 in the coupling phase diagram are given by
\begin{align}
&\left(g_{1}^{\text{(3)}}\big|_{\text{EP3}},\, g_{2}^{\text{(3)}}\big|_{\text{EP3}}\right) \nonumber \\
& \qquad = \left(\sqrt{\frac{4(2\gamma_{-}+\kappa_{-})^{3}}{27(\gamma_{-}+\kappa_{-})}},\, \sqrt{\frac{4(2\kappa_{-}+\gamma_{-})^{3}}{27(\gamma_{-}+\kappa_{-})}}\right). 
\label{eq:3modeEP3}
\end{align}
Note that the preceding analysis does not reveal exact nature of coupling between the modes, or identify which modes are strongly coupled in the white region of Fig.~\ref{Fig:OpenLoop}(b); we can only ascertain that there exists at least one pair of modes that is strongly coupled for ${(g_1^{(3)} > \gamma_{-}) \cap (g_2^{(3)}>\kappa_{-})}$. For detailed analysis and explicit expressions for eigenvalues in a three-mode system, we refer the reader to appendix \ref{App:Evalanalyis}.
%
%
\subsection{$N=4$ case}
%
We consider a four-mode system described by a mode matrix of the form,
\begin{equation}
\mathbb{M}^{(4)} = 
\left(
\begin{array}{cccc}
-\kappa_1/2 & -ig_{1}^{(4)} & 0 & 0 \\
-ig_{1}^{(4)} & -\kappa_{2}/2 & -ig_{2}^{(4)} & 0 \\
0 & -ig_{2}^{(4)} & -\kappa_{3}/2 & -ig_{1}^{(4)} \\
0 & 0 & -ig_{1}^{(4)} & -\kappa_{4}/2
\end{array} 
\right).
\label{Eq:ModeMat4}
\end{equation}
As before, we consider resonant driving, with $g_{12} = g_{34} = g_{1}^{(4)}$, $g_{23} = g_{2}^{(4)}$ where $g_{j}^{(\rm{4})}\in\mathbb{R_{>\text{0}}}\,\forall j$. Besides allowing us to restrict our analysis to a 2D phase diagram, this pattern of alternating couplings is of relevance to interesting physical models, such as Su-Schrieffer-Heeger (SSH) model \cite{Su1979,Heeger1988} describing hopping of spinless fermions on a 1D lattice \cite{Li2014}.
\par
Proceeding as in the case of three modes, we obtain the EP2s of the four-mode system as a function of coherent couplings $g_{1,2}^{(4)}$ for fixed decay rates $\kappa_{j}$ (see appendix \ref{App:Evalanalyis} for details). As shown in Fig.~\ref{Fig:OpenLoop}(c), we now obtain three EP2 curves with three intercepts on the axes. The intercepts on the x-axis correspond to $g_{1}^{(4)} = |\kappa_{2} - \kappa_{1}|/4$ and $g_{1}^{(4)} = |\kappa_{4} - \kappa_{3}|/4$, setting the strong coupling thresholds for decoupled $\{a_1,a_2\}$ and $\{a_3,a_4\}$ subsystems respectively. Similarly, the y-intercept corresponds to $g_{2}^{(4)} = |\kappa_{3} - \kappa_{2}|/4$, the strong coupling threshold for decoupled $\{a_2, a_3\}$ subsystem. However, as in the case for three-modes, in regions sufficiently distant from the axes, the identity of the modes that are strongly coupled remains ambiguous.
%
%
\section{Strong coupling analysis based on eigenvectors}
\label{Sec:evecs}
%
As is evident from the discussion in the previous section, while EPs provide a clear separation of weak and strong coupling regimes, they fail to identify the physical modes that span the strongly-coupled subspace in a multi-mode ($N>2$) system past EP$N$ [white region of Figs.~\ref{Fig:OpenLoop}(b) and (c)]. In this section, we introduce a new method based on 2D planar projections of eigenvectors which provides a universal way to detect $N$-way hybridization, complete with an identification of the strongly-coupled subspace, in a multi-mode open system.
\par
We begin with a simple two-mode example to illustrate the behavior of eigenvectors in weak and strong coupling regimes. To this end, we consider the amplitudes of (normalized) left eigenvectors $V^{\alpha}$, 
\begin{eqnarray}
    |V^{\alpha}|\equiv\left[\;|(V^{\alpha}, a_1)|, |(V^{\alpha}, a_2)|\;\right]
\end{eqnarray}
where $a_{j}\equiv [1_{j}]$ denotes a basis vector with unity as the $j^{\rm th}$ physical mode and zero for every other entry, and (u,v) represents the vector inner product. The vector $|V^{\alpha}|$ can be thought of as a ``participation ratio vector'' since each entry denotes the participation ratio of physical mode $a_i$ in the eigenmode. For $g^{(2)}=0$, $|V^1|= a_{1} = [1, 0]$ and $|V^2|= a_{2} = [0, 1]$; hence $(|V^1|, |V^2|)=0$ since $a_1$ and $a_2$ are orthogonal basis vectors. Throughout the weak coupling regime $|g^{(2)}|<|g^{(2)}_{\text{EP2}}|$, $0\leq(|V^1|,|V^2|)<1$.
On the other hand, in the strong coupling regime, $|g^{(2)}|>|g^{(2)}_{\text{EP2}}|$, $(|V^1|,|V^2|)=1$, implying that $|V^{1}|$ and $|V^{2}|$ are parallel. While the above example shows how distinct nature of eigenvectors, without any knowledge of the eigenvalues, can provide a sufficient means for distinguishing the different regimes of coupling, one should be wary of naively extending the two-mode intuition to a multi-mode system. For instance, one potential pitfall is to assume identical participation ratios post hybridization into supermodes as a criterion for mode indistinguishability in the strong-coupling regime.  While for a two-mode system, the participation ratios indeed become identical at EP2, i.e. $|V^{1,2}_{a_1}|_{g=g^{(2)}_{EP2}}=|V^{1,2}_{a_2}|_{g=g^{(2)}_{EP2}}=1/\sqrt{2}$, for $N>2$ systems $|V^{\alpha}_{a_j}|\neq1/\sqrt{N}$ at (or beyond) EP$N$ in general. For instance, at EP3 for $N=3$ [Fig.~\ref{Fig:OpenLoop}(b)], for each of the three eigenvectors $|V^{1,2,3}|=[0.42,0.71,0.57]$. In other words, $N$-way strong coupling does not guarantee equal participation of the modes in a generic $N$-mode system.
\par
\begin{figure}[t]
\centering
\begin{minipage}[b]{0.9\linewidth}
\includegraphics[width=\textwidth]{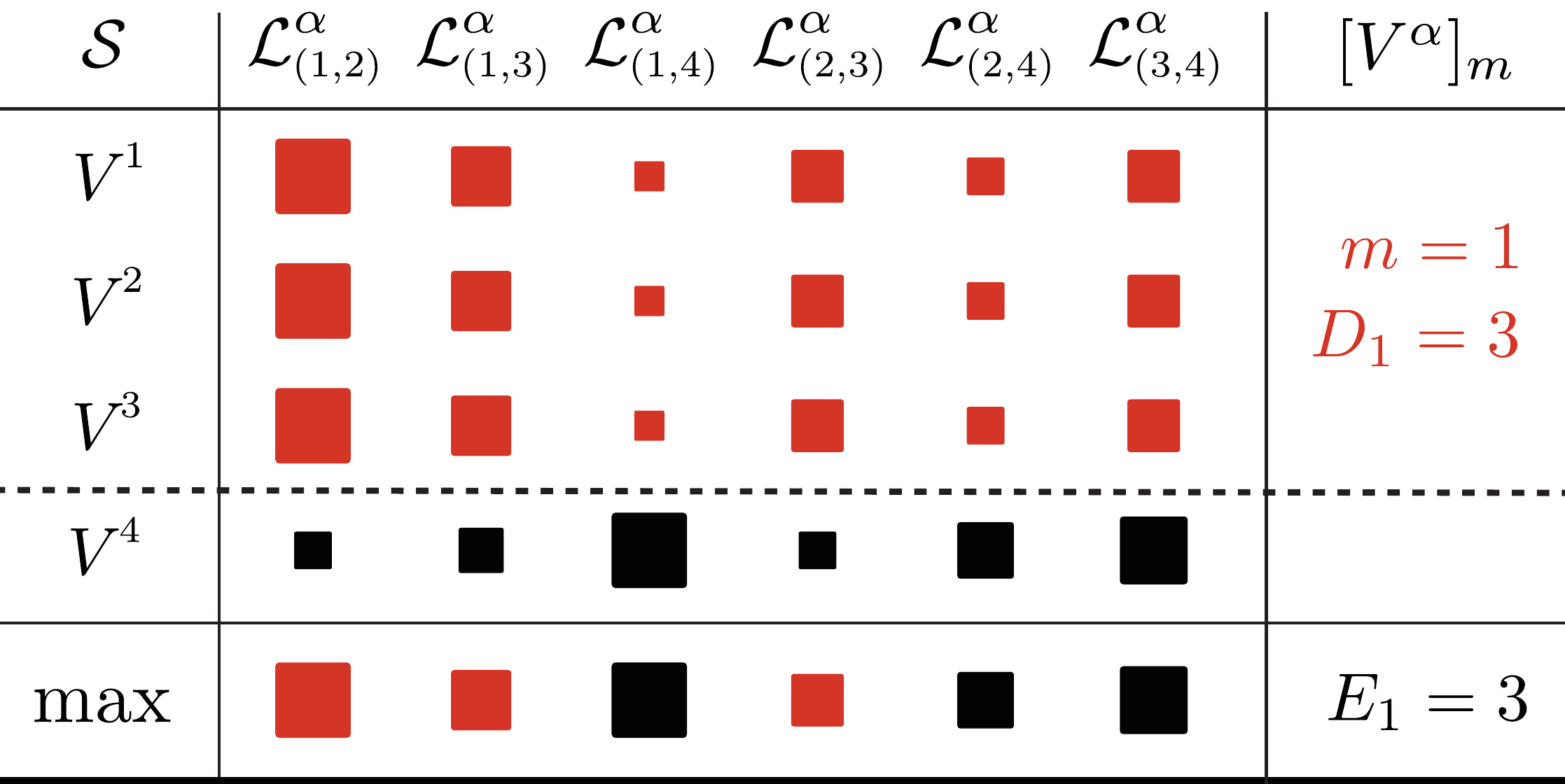}
\label{Fig:EP31}
\end{minipage}\hfill%
\vspace{-5pt}
\begin{minipage}[t]{0.7\linewidth}
\includegraphics[width=\textwidth]{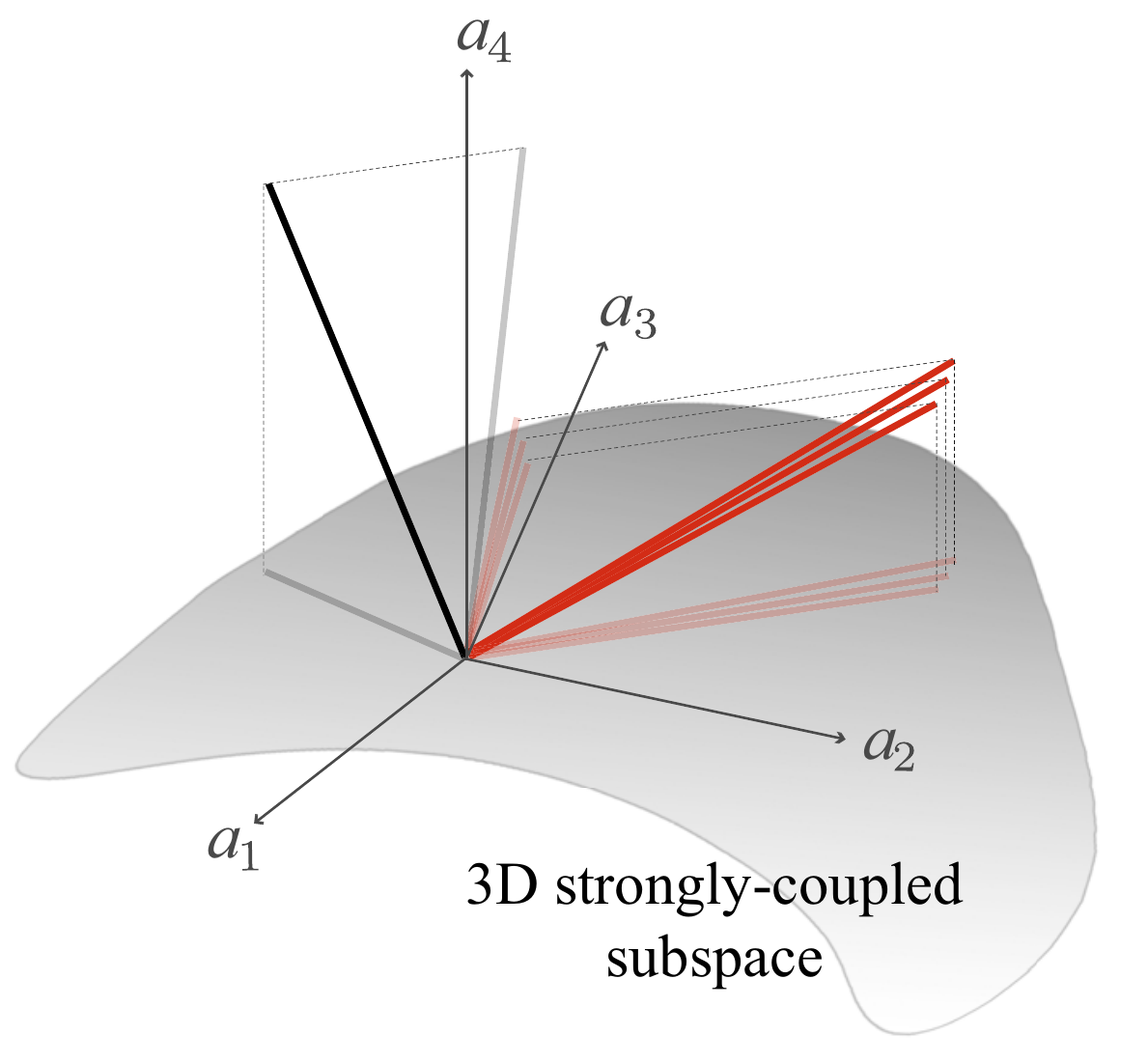}
\label{Fig:EP31}
\end{minipage}
\vspace{-10pt}
\caption{(\emph{Top panel}) Configuration of eigenvectors corresponding to a realization of EP3 in $N=4$ system. The size of the squares denote the magnitude of planar projections with red denoting those for strongly-coupled modes and black denoting those for the weakly coupled mode. (\emph{Bottom panel}) Geometric visualization of EP3 in $N=4$ system. Solid rays represent the eigenvectors while faint rays represent their corresponding projections. For clarity of presentation only the projections in $\{a_{1},a_{2}\}$ and $\{a_{3},a_{4}\}$ planes are shown. The resultant EP3 creates a strongly-coupled 3D subspace spanned by $\{a_{1},a_{2},a_{3}\}$, represented as a grey hyperplane.}
\label{Fig:EP3}
\end{figure}
We now introduce the full procedure based exclusively on eigenvector analysis, which reliably diagnoses strong coupling in a general $N$-mode system with bilinear interactions. Note that the modes under consideration may or may not share direct physical coupling, for instance, including non-nearest neighbor sites in the systems considered in Sec.~\ref{Sec:evals}. 
\begin{itemize}
\item Consider the multiset of left eigenvectors of the mode matrix, $\mathcal{S} = \{V^{\alpha}\; |\;V^{\alpha}\mathbb{M}^{(N)}=V^{\alpha}\lambda^{\alpha}\}$. 
\item Define the 2-norm of each eigenvector $V^{\alpha}$, projected onto a 2D subspace spanned by $\{a_j,a_k\}$ as,
\begin{eqnarray}
\hspace{1cm}\mathcal{L}^{\alpha}_{(j,k)} = \left(\big|(V^{\alpha}, a_j)\big|^{2} + \big|(V^{\alpha}, a_k)\big|^{2}\right)^{1/2}.
\label{eq:normsquared}
\end{eqnarray}
\item Partition $\mathcal{S}$ into $m$-equivalence classes $[V^{\alpha}]_{m}$, each consisting of a set of eigenvectors with equal 2-norms for all $\prescript{N}{}{\mathbf{C}}_{2} $ projections, i.e.
\begin{eqnarray}
[V^{\alpha}]_{m} = \{V^{\beta} \in \mathcal{S}\;| \;V^{\beta} \sim V^{\alpha}\},
\label{Eq:order}
\end{eqnarray}
if $\mathcal{L}^{\alpha}_{(j,k)} = \mathcal{L}^{\beta}_{(j,k)}, \forall (j,k) \in [1,N]$. The size of each equivalence class defines the \emph{coupling depth}, ${D_m= |[V^{\alpha}]_{m}|\leq N}$, for each $D_m$-dimensional strongly-coupled subspace of the $N$-mode system.
\item If $|[V^{\alpha}]_{m}|= 1 \; \forall \; m$, this implies that all modes are weakly coupled. 
\item Two modes $a_j$ and $a_k$ are strongly coupled, if and only if,
\begin{eqnarray}
    \hspace{1cm} \mathcal{L}^{\alpha}_{(j,k)} > \mathcal{L}^{\alpha'}_{(j,k)}
  \; \;\forall \; V^{\alpha'}\not\in [V^{\alpha}]_{m}.
\label{Eq:metric}
\end{eqnarray}
Using the above inequality, construct a set $\mathcal{E}_{m}$
\begin{eqnarray}
    \hspace{1cm} \mathcal{E}_{m} = \left\{ (j,k); {j<k} \;|\; \mathcal{L}^{\alpha}_{(j,k)} > \mathcal{L}^{\alpha'}_{(j,k)},  \; \forall \; V^{\alpha'} \notin [V^{\alpha}]_{m} \right\}, \nonumber\\
\label{Eq:connectivity}
\end{eqnarray}
whose size defines the \emph{connectivity} of the subsystem, $E_{m}=|\mathcal{E}_{m}|$. Connectivity represents the number of pairs of physical modes $(j,k)$ that are hybridized, i.e., each pair of modes in $\mathcal{E}_{m}$ indexes a 2D subspace in the $N$-dimensional (physical) mode space. 
\item The connectivity $E_{m}$ is distinct from the depth $D_m$ and, in general, $E_{m} \geq D_m -1$. If all pairs in $\mathcal{E}_{m}$ form a fully-connected closed set, then $E_{m} =\prescript{D_m}{}{\mathbf{C}}_{2}$ and the subspace supports an EP$D_{m}$. 

Note that this implies that for $D_m=N$, Eqs.~(\ref{Eq:order})-(\ref{Eq:metric}) recover the condition of an EP$N$, i.e., coalescence of all eigenvectors of the system signifying the manifestation of $N$-way strong coupling in an $N$-mode system. 
\end{itemize}
The criterion prescribed in Eq.~(\ref{Eq:metric}), which is the key result of this paper, lends itself to a helpful geometric visualization depicted in Fig.~\ref{Fig:EP3}: strongly-coupled subspaces manifest as hyperplanes making small angles with the equivalent eigenvectors thus making the corresponding projections larger, while weakly coupled subspaces make large angles leading to small projections. 
\par
\begin{figure*}[t!]
\centering
\includegraphics[width=0.9\textwidth]{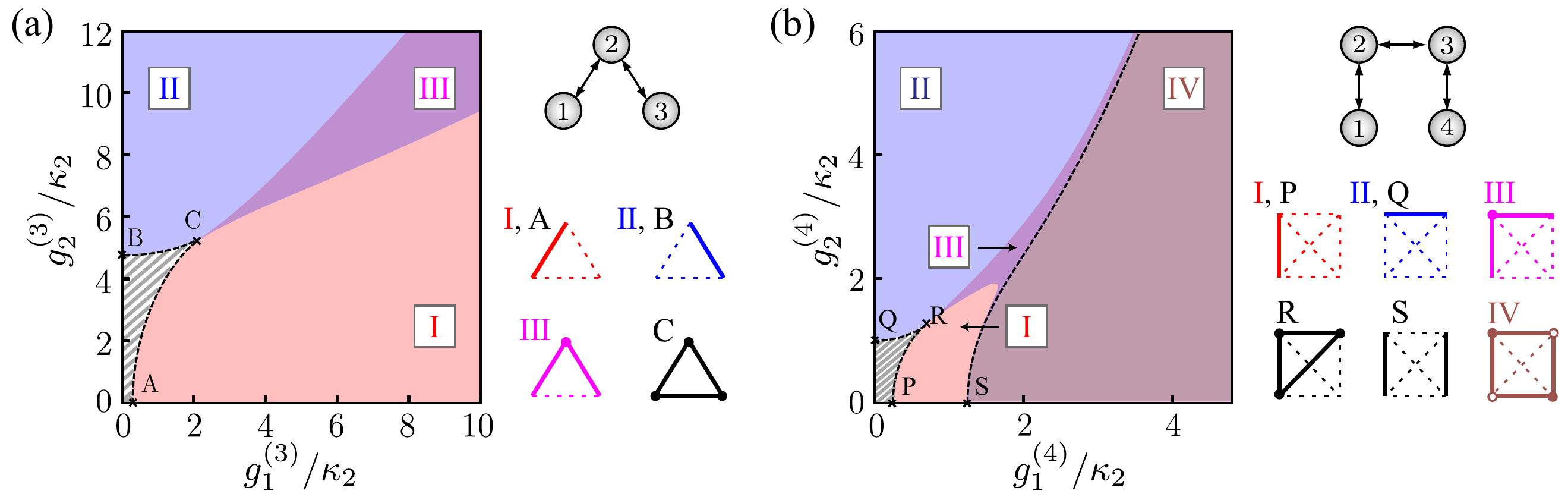}
\vspace{-10pt}
 \caption{Pairwise strong coupling regions calculated using Eq.~($\ref{Eq:metric}$) for (a) $N=3$ and (b) $N=4$ modes, depicted as function of respective coupling strengths. The decay rates used in each case were the same as those reported in Fig.~\ref{Fig:OpenLoop}. The hatched region in each plot depicts the weak coupling region where the inequality is not satisfied for any pair of modes. The boundaries of the regions predicted by Eq.~($\ref{Eq:metric}$) coincide with the EP2 curves obtained from eigenvalues, which are reproduced here in dashed-black for easy reference [cf. Fig.~\ref{Fig:OpenLoop}]. Along with each coupling map, corresponding edge graphs show mode connectivity in each region and at EPs inferred from eigenvector-based projection, with connections between strongly-coupled (weakly-coupled) modes represented with solid (dashed) edges. Connected solid edges, resulting from the same equivalence class, are shown with filled circles. Disconnected solid edges denote presence of distinct equivalence classes of dressed states, shown with empty circles for adjacent edges; for example, point S and region IV in (b).}
 \label{Fig:OLPairwiseSCRegimes}
\end{figure*}

We now apply this procedure to the three-mode and four-mode systems examined in Sec.~\ref{Sec:evals}. Figure~\ref{Fig:OLPairwiseSCRegimes}(a) depicts the regions where Eq.~(\ref{Eq:metric}) holds true for each pair of modes in ${N=3}$ open system. For instance, in region I (red), ${D=2, E=1}$ with two identical eigenvectors such that ${\mathcal{L}^{1}_{(1,2)} = \mathcal{L}^{2}_{(1,2)}  > \mathcal{L}^{3}_{(1,2)}}$ while ${\mathcal{L}^{1}_{(2,3)}  = \mathcal{L}^{2}_{(2,3)} < \mathcal{L}^{3}_{(2,3)}}$ and ${\mathcal{L}^{1}_{(1,3)} = \mathcal{L}^{2}_{(1,3)} < \mathcal{L}^{3}_{(1,3)}}$,  identifying this region as regime of pairwise strong coupling for modes $\{a_{1},a_{2}\}$.
Similarly, in region II (blue), ${D=2, E=1}$ with ${\mathcal{L}^{1}_{(1,2)} = \mathcal{L}^{2}_{(1,2)} < \mathcal{L}^{3}_{(1,2)}}$, ${\mathcal{L}^{1}_{(2,3)}  = \mathcal{L}^{2}_{(2,3)} > \mathcal{L}^{3}_{(2,3)}}$, ${\mathcal{L}^{1}_{(1,3)} = \mathcal{L}^{2}_{(1,3)} < \mathcal{L}^{3}_{(1,3)}}$, identifying pairwise strong coupling between modes $\{a_{2},a_{3}\}$ in this region. Furthermore, boundaries of regions I and II delineate weak and strong coupling regimes based on eigenvector analysis, which on comparison with Fig.~\ref{Fig:OpenLoop}(a) are in quantitative agreement with the EP2 curves obtained from eigenvalue analysis. More interestingly, our analysis identifies a region III (purple) where regions I and II overlap, i.e. $D=2, E=2$, implying simultaneous pairwise strong coupling for two pairs of modes, $\{a_1,a_2\}$ and $\{a_2,a_3\}$. Note that this does not imply that all three modes are strongly coupled in region III, because $\{a_{1},a_{3}\}$ remain weakly coupled since $\mathcal{L}^{1}_{(1,3)} = \mathcal{L}^{2}_{(1,3)} < \mathcal{L}^{3}_{(1,3)}$ remains true in all the colored regions. In fact, the only point in parameter space $(g_1^{(3)}, g_2^{(3)})$ that supports 3-way strong coupling is point B; here ${\mathcal{L}^{1}_{(j,k)} = \mathcal{L}^{2}_{(j,k)} = \mathcal{L}^{3}_{(j,k)}}$ where $(j,k) \in [1,3]$. It is worth noting that this exactly corresponds to the EP3 shown in Fig.~\ref{Fig:OpenLoop}(a). 
\par
The coupling phase diagram shown for ${N=4}$ in  Fig.~\ref{Fig:OLPairwiseSCRegimes}(b) is expectedly more involved. In total there are 6 pairs for which we check Eq. (\ref{Eq:metric}), and find in
\begin{eqnarray*}
& & \text{region I:} \; D=2, E=1,\; \text{since}\\
& & \qquad \qquad \qquad \mathcal{L}^{1}_{(1,2)} = \mathcal{L}^{2}_{(1,2)}  > \mathcal{L}^{3,4}_{(1,2)},\\
& &\text{region II:} \; D=2, E=1,\; \text{since}\\
& & \qquad \qquad \qquad \mathcal{L}^{1}_{(2,3)} = \mathcal{L}^{2}_{(2,3)}  > \mathcal{L}^{3,4}_{(2,3)},\\
& & \text{region III:}\; D=2, E=2, \; \text{since}\\
& & \; \bigcap_{\substack{(j,k) \in \mathcal{E}\\ \mathcal{E}=\{(1,2), (2,3)\}}} \mathcal{L}^{1}_{(j,k)} = \mathcal{L}^{2}_{(j,k)}  > \mathcal{L}^{3,4}_{(j,k)}.
\end{eqnarray*}
Here, for brevity, we report only the pairs of modes that satisfy Eq.~(\ref{Eq:metric}) for strong-coupling in the respective regions. In each region, for pairs $(j,k) \not\in \mathcal{E}$, 
\begin{eqnarray*}
\bigcap_{\substack{(j,k) \not\in \mathcal{E}}} \mathcal{L}^{1}_{(j,k)} = \mathcal{L}^{2}_{(j,k)}  < \mathcal{L}^{3,4}_{(j,k)}.
\end{eqnarray*}
Note that in all the regions only 2-way strong coupling, i.e. ${D=2}$, is realized. Though more than one pair of modes are strongly coupled in regions III and IV, 3- or 4-way strong-coupling is not realized in these regions since $\{a_1,a_3\}$ and $\{a_2,a_4\}$ are diagnosed as weakly coupled, violating the condition of full connectivity necessary for realizing higher coupling depth $D$. The transition from $S$ to region IV is particularly noteworthy, even though it entails no change in the coupling depth. Both these regions support two distinct equivalence classes of eigenvectors, each consisting of a pair of identical vectors i.e. $D_{1}=D_{2}=2$. However, while at $S$ these support two decoupled 2D subspaces with $E_{1}=E_{2}=1$ since
\begin{eqnarray*}
& & \mathcal{L}^{1}_{(1,2)} = \mathcal{L}^{2}_{(1,2)}  > \mathcal{L}^{3,4}_{(1,2)} \\
\text{and} & & \; \mathcal{L}^{3}_{(3,4)} = \mathcal{L}^{4}_{(3,4)}  > \mathcal{L}^{1,2}_{(3,4)},
\end{eqnarray*}
in region IV, even a very weak coupling $g_{2}$ couples these 2D subspaces leading to $E_{1}=E_{2}=2$  since
\begin{eqnarray*}
& & \; \bigcap_{\substack{(j,k) \in \mathcal{E}_{1}\\ \mathcal{E}_{1} = \{(1,2),(2,3)\}}} \mathcal{L}^{1}_{(j,k)} = \mathcal{L}^{2}_{(j,k)}  > \mathcal{L}^{3,4}_{(j,k)}\\
\text{and} & & \; \bigcap_{\substack{(j,k) \in \mathcal{E}_{2}\\ \mathcal{E}_{2}= \{(1,4),(3,4)\}}} \mathcal{L}^{3}_{(j,k)} = \mathcal{L}^{4}_{(j,k)}  > \mathcal{L}^{1,2}_{(j,k)}.
\end{eqnarray*}
Thus each equivalence class of vectors contributes a pair of adjacent edges that combine to realize four-mode hybridized states, as indicated by the respective edge diagram in Fig.~\ref{Fig:OLPairwiseSCRegimes}(b). This is an open-system analogue of the superexchange interaction describing electron transfer in strongly-correlated systems, where two strongly-correlated electronic states can hybridize through a weakly-correlated state \cite{Kanamori1959}. This instance shows how information about connectivity between physical modes of a multi-mode system can reveal physics beyond that provided by coupling depth.
\par
At point R in Fig.~\ref{Fig:OLPairwiseSCRegimes}(b), $D=3, E=3$ with
\begin{eqnarray*}
\bigcap_{\substack{(j,k) \in \mathcal{E}\\ \mathcal{E} = \{(1,2), (2,3), (1,3)\}}}\mathcal{L}^{1}_{(j,k)} = \mathcal{L}^{2}_{(j,k)} =  \mathcal{L}^{3}_{(j,k)}  > \mathcal{L}^{4}_{(j,k)},
\end{eqnarray*}
while $\mathcal{L}^{1,2,3}_{(j,k)} < \mathcal{L}^{4}_{(j,k)}$  for $(j,k) = (1,4), (2,4), (3,4)$. This diagnoses 3-way strong coupling in $\{a_1,a_2,a_3\}$ subsystem which, as in the case of $N=3$, coincides with EP3 for this system predicted by eigenvalues [c.f. Fig.~\ref{Fig:OpenLoop}(b)]. Further, the boundaries of different regions identified using eigenvector projections correspond exactly to the EP2 curves of Fig.~\ref{Fig:OpenLoop}(b) with the weak-coupling regime corresponding to the region where Eq. (\ref{Eq:metric}) is violated for every pair of modes. Thus in addition to correctly predicting coordinates of EPs in parameter space, eigenvectors also provide information about which modes of system hybridize at each EP$N$. 
\par
We emphasize that the preceding analysis makes exclusive use of eigenvectors, without invoking eigenvalues of the mode matrix. The proposed inequality in Eq.~(\ref{Eq:metric}) relies on 2D projections of $N$-dimensional eigenvectors, which indicates that analyzing pairwise-coupled subspaces is sufficient to diagnose \emph{arbitrary} coupling depth in open systems with bilinear interactions. Further, eigenvector analysis supersedes the information obtained from usual EP physics unraveled by eigenvalues, by providing means to identify physical modes defining the strongly-coupled subsystems in a multi-mode system. 
%
%
\section{Application: Dissipation-engineered cooling}
\label{Sec:cooling}
\begin{figure}[t!]
\centering
\includegraphics[width=\columnwidth]{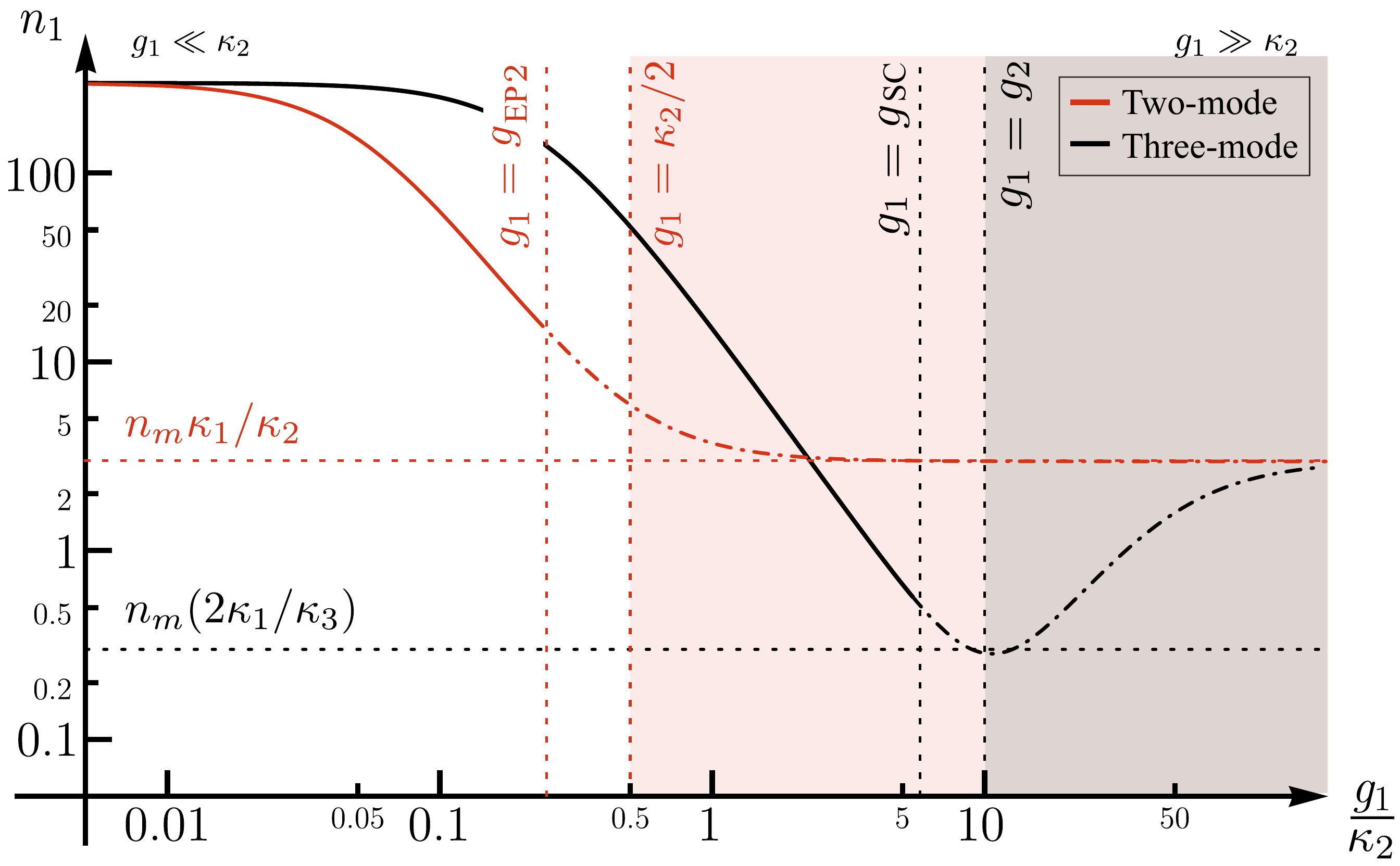}
\vspace{-10pt}
 \caption{Phonon population versus coupling strength for the two-mode ($g_{2}=0$) and three-mode ($g_2/\kappa_{2} = 10, \;\mathcal{C}_{2} =20$) systems with $n_{m} =300,\; n_{0}=n_{a}=0.1$, calculated for the same decay rates as used in Fig. \ref{Fig:OpenLoop}(b). The red (gray) region represents the decay-dominated regime for the two-mode (three-mode) system, calculated using expression for phonon population in Eq.~(\ref{Eq:twomodecooling}) [(\ref{Eq:threemodecooling})]. The horizontal dashed lines correspond to the minimum achievable population, while the vertical dashed lines correspond to critical $g_{1}$ values indicating the onset of strong coupling and the decay-dominated regimes. For each system, the population is shown with dashed-dotted curves in the regime where mechanical mode is hybridized with optics. The onset of this region for the two-mode system ($g_{\rm EP2}$) is estimated using the eigenvalues [cf.~Fig.~\ref{Fig:OpenLoop}(b)], while for the three-mode system ($g_{\rm SC}$) it is estimated using the eigenvector projection-based method introduced in Sec.~\ref{Sec:evecs} [cf.~Fig.~\ref{Fig:OLPairwiseSCRegimes}].}
 \label{Fig:Cooling}
 \end{figure}

%
%
In this section, we elucidate the physical implications of the eigenvector-based strong coupling diagnostic by applying it to the problem of quantum ground state cooling. Cooling quantum systems is a mainstay in many quantum information platforms where a mode (or qubit) needs to be prepared in its ground state (or `reset'). For instance, in conventional optomechanical platforms, a hot mechanical oscillator ($a_{1}$) is parametrically coupled to a cold optical resonator ($a_{2}$) that acts as an engineered reservoir. On modulating the coupling at the difference frequency of the two modes, the mechanical mode is cooled by shuttling excitations to the optical mode, which decays at a sufficiently fast rate to beat the (equally likely) reverse conversion process. The resultant phonon population in the steady state for the resolved sideband regime is \cite{Aspelmeyer2014}
\begin{eqnarray}
n_{1}^{(2)} & = & n_{m}\frac{1 + \kappa_{1}/\kappa_{2}(1+\mathcal{C}_{1})}{(1+\kappa_{1}/\kappa_{2})(1+\mathcal{C}_{1})} + n_{o}\frac{\mathcal{C}_{1}}{(1+\kappa_{1}/\kappa_{2})(1+\mathcal{C}_{1})} \nonumber\\
&\approx& n_{m}\left(\kappa_{1}/\kappa_{2} + 1/\mathcal{C}_{1}\right) + n_{o},
\label{Eq:twomodecooling}
\end{eqnarray}
where $\kappa_{1,2}$ denote the decay rates associated with the mechanical and optical modes, $n_{m}$ and $n_{o}$ denote their respective thermal populations in the absence of coupling, and the coupling strength $g_{1}$ is parametrized in terms of cooperativity $\mathcal{C}_{1}= 4g_{1}^{2}/\kappa_{1}\kappa_{2}$. From the simplified expression obtained in the limit of large cooperativity $\mathcal{C}_{1} \gg 1$ and the typical decay hierarchy $\kappa_{1}/\kappa_{2} \ll 1$, we can identify two distinct regimes of operation: (i) cooperativity-dominated, or $\kappa_{1}/\kappa_{2} \ll 1/\mathcal{C}_{1}$, and (ii) decay-dominated regimes, or $\kappa_{1}/\kappa_{2} \gg 1/\mathcal{C}_{1}$. As is evident from the red curve in Fig.~\ref{Fig:Cooling}, the mechanical mode experiences active cooling as long as the system is the cooperativity-dominated regime. For coupling strengths $g_{1}/\kappa_{2} > 1$ the population becomes independent of $g_{1}$ and saturates to the steady state value determined by bare decay rates $n_{1, {\rm min}}^{(2)} = n_{m}(\kappa_{1}/\kappa_{2})$. This crossover into dissipation-dominated regime is intimately related to the onset of strong coupling and hybridization of the mechanical and optical modes at $g_{1} = g_{\rm EP2}$, which eventually manifests as saturation of phonon population \cite{Dobrindt2008}. 
\par
The threshold for this crossover into strong coupling can be modified by coupling the mechanical mode to a more complex bath. The minimal system to implement this is the three-mode system considered in Sec.~\ref{Sec:evals}, where a second optical mode $a_{3}$ is introduced as an additional auxiliary reservoir with no direct coupling to the mechanics $a_{1}$. The goal is to delimit the regime where the target system ($a_{1}$) remains weakly coupled with the system of engineered reservoir modes ($a_{2},a_{3}$), in order to extend the cooperativity-dominated regime for cooling. Based on the coupling phase diagram of Fig.~\ref{Fig:OLPairwiseSCRegimes}(a), this may be achieved if we choose to operate in region II where $\{a_{1},a_{2}\}$ and $\{a_{1},a_{3}\}$ subsystems remain weakly coupled, while optical baths $a_{1}$ and $a_{2}$ hybridize to form supermodes. 
\par
To demonstrate this, we follow the same procedure as for the two-mode case and calculate the phonon population as a function of coupling of the mechanics to the system of optical cavities $g_{1}$.  
To gain some intuition of the modified strong coupling threshold, we first treat the auxiliary optical mode $a_{3}$ as quasi-static, ${\kappa_{3}\gg{\rm max}\{\kappa_{1},\kappa_{2}\}}$, and use its steady state solution,
\begin{eqnarray}
    a_{3} = \frac{2}{\kappa_{3}}\left(-ig_{2} a_{2} + \sqrt{\kappa_{3}}a_{3}^{\rm in}\right),
\end{eqnarray}
to solve for dynamics of the reduced two-mode system $\{a_{1},a_{2}\}$. In this limit, the mechanical mode can be viewed as being coupled to a single optical mode $a_{2}$ with a modified decay rate $\kappa_{2}^{\text{eff}} = \kappa_{2}\left(1+\mathcal{C}_{2}\right)$, and a concomitant input noise $a_{2}^{\rm in, \text{eff}} = a_{2}^{\rm in} - i\sqrt{\mathcal{C}_{2}} a_{3}^{\rm in}$, where $\mathcal{C}_2 = 4g_{2}^2/(\kappa_2\kappa_3)$ denotes the cooperativity for the optical subsystem. Following standard procedure, we find the phonon population for this effective two-mode system as
\begin{align}
    n_{1}^{(2),\text{eff}}  &= n_m \frac{(1 + \mathcal{C}_{2}) +\kappa_{1}/\kappa_{2}^{\text{eff}}(1+\mathcal{C}_{1}+\mathcal{C}_{2})}
    {(1 + \kappa_{1}/\kappa_{2}^{\text{eff}})(1+\mathcal{C}_{1}+\mathcal{C}_{2})} \nonumber \\
    & \quad + n_{o}\frac{\mathcal{C}_{1}/(1+\mathcal{C}_{2})}
    {(1+ \kappa_{1}/\kappa_{2}^{\text{eff}}) (1+\mathcal{C}_{1}+\mathcal{C}_{2})} \nonumber \\
    & \quad + n_{a}\frac{\mathcal{C}_{1}\mathcal{C}_{2}/(1 + \mathcal{C}_{2})}
    {(1 + \kappa_{1}/\kappa_{2}^{\text{eff}})(1+\mathcal{C}_{1}+\mathcal{C}_{2})},
\label{Eq:threemodAEecooling}
\end{align}
where $n_{m}$, $n_{o}$ and $n_{a}$ denote the intrinsic populations of the mechanical mode and optical modes in the absence of couplings. In the limit of ${\mathcal{C}_{1}\rightarrow\infty}$, $n_{1, {\rm min}}^{(2),\text{eff}}= n_{m}(\kappa_{1}/\kappa_{2}^{\text{eff}})$ analogous to the conventional two-mode system. This simple analysis indicates that in the presence of an additional decay channel presented by the auxiliary mode $a_{3}$, strong-coupling threshold may be realized at a higher value corresponding to the high effective decay rate presented by the bath modes. However, an adiabatic elimination of $a_{3}$ strictly holds true for $\mathcal{C}_{2} \leq 1$. In order to obtain phonon population for strong coupling between optical modes --- which is the regime of interest for operating in region II of Fig.~\ref{Fig:OLPairwiseSCRegimes}(a) --- we perform the calculation for the full three-mode system including the dynamics of the auxiliary optical mode. For full details of this calculation, we refer the reader to appendix \ref{App:full3mode}; here we present the simplified expression for phonon population, obtained in the limit of large cooperativities ($\mathcal{C}_{1,2} \gg 1$) and for the decay hierarchy $\kappa_{3} \gg \kappa_{2} > \kappa_{1}$,
\begin{eqnarray}
    n_{1}^{(3)} & \approx & n_{m} \left( \frac{\mathcal{C}_{2}}{\mathcal{C}_{1} + \mathcal{C}_{2}}  + \frac{\kappa_{1}^{2}}{\kappa_{3}^{2}}\frac{\mathcal{C}_{1}}{\mathcal{C}_{2}}\right)
    + n_{o} \left(\frac{\mathcal{C}_{1}}{\mathcal{C}_{1} + (\kappa_{3}^{2}/\kappa_{1}\kappa_{2}) \mathcal{C}_{2}}\right)  \nonumber\\ 
   & & + \; n_{a} \left(\frac{(\kappa_{3}^{2}/\kappa_{1}\kappa_{2}) \mathcal{C}_{2}}{\mathcal{C}_{1} + (\kappa_{3}^{2}/\kappa_{1}\kappa_{2}) \mathcal{C}_{2}}\right).
\label{Eq:threemodecooling}
\end{eqnarray}
Following a similar line of logic as for the two-mode case, we can distinguish the cooperativity-dominated regime ($\mathcal{C}_{2}/\mathcal{C}_{1} \gg \kappa_{1}/\kappa_{3}$) from the decay-dominated regime ($\mathcal{C}_{2}/\mathcal{C}_{1} \ll \kappa_{1}/\kappa_{3}$) of operation by analyzing the coefficient of the $n_{m}$ term. Interestingly, the crossover between these two regimes is realized when the two couplings are balanced, i.e. $g_{1} = g_{2}$. As shown by the result of the full calculation (black curve in Fig.~\ref{Fig:Cooling}), this is also the point where the lowest phonon population is achieved with the floor, $n_{1, {\rm min}}^{(3)} = 2 n_{m} \kappa_{1}/\kappa_{3}$, determined solely by the decay rates. This indicates that while the quantum correlations of reservoir modes enhance cooling, eventually strong coupling effects lead to a resurgence observed for large values of $g_{1}$. 
\par
Note that, unlike the two-mode case where the real and imaginary parts of eigenvalues exhibit a bifurcation as the system crosses EP2, the eigenvalues of $\mathbb{M}^{(3)}$ show no characteristic signature as this crossover is approached [see Fig.~\ref{Fig:OpenLoop}(b)]. However, we can evaluate a threshold value for $g_{1}$, given a value of $g_{2}$, using the metric proposed in Eq.~(\ref{Eq:metric}), below which mechanical mode remains weakly coupled. This value of $g_{1} = g_{\rm SC}$ corresponds to the intersection of the line $g_{2}/\kappa_{2} =10$ with the boundary of regions II and III in Fig.~\ref{Fig:OLPairwiseSCRegimes}(a). Notably, as shown in Fig.~\ref{Fig:Cooling}, the predicted value of $g_{\rm SC}$ is consistent with the fact that hybridization of the modes acts as a precursor for population saturation,  and beyond this point the cooling is progressively impeded with increase in coupling. Thus, eigenvector-based analysis is able to detect the transition from weak-to-strong coupling in dissipation-engineered systems, which cannot be discerned by analyzing eigenvalues. 
%
%
\section{Conclusions}
\label{Sec:conclusions}
%
In conclusion, we have introduced a new method to diagnose strong coupling in a multi-mode open system with bilinear interactions. The proposed method is based entirely on eigenvectors of the matrix describing the coupling and local decay rates of the modes. In addition to delineating the regions of weak and strong coupling, it allows a means to identify the physical subsystems that undergo hybridization in different regions of the coupling landscape and shows how different connectivity configurations can be present while maintaining a fixed coupling depth. This indicates that detailed information about both connectivity and coupling depth is essential for a full characterization of hybridized states/subsystems in strongly-coupled systems. We present sideband cooling in a multi-mode optomechanical system as an example to show how this method can reveal the crossover of the target oscillator from cooperativity-dominated dynamics to decay-dominated dynamics in the presence of a strongly-hybridized optical reservoir. Thus using eigenvectors to characterize open system dynamics, which cannot be detected by EPs, can present new opportunities for dissipation engineering where, by construction (or design!), only a subsystem is accessible for control and measurement.
\par
Remarkably, the proposed method shows how tiling only pairwise hybridized modes can detect exceptional points of arbitrary order (at least for bilinear interactions). This is strikingly reminiscent of dimensional reduction methods used for feature analysis of multi-dimensional data. The current work thus just scratches the surface in adapting sophisticated data analytics tools to resolve challenging problems in many-body open systems. For instance, leveraging connections to statistical techniques such as projection pursuit, the eigenvector-based method presented here may be generalized to different coupling topologies,  PT symmetric systems \cite{Bender1998,El-Ganainy2018} and systems with gain \cite{Miri2019}, and even nonlinear couplings. Finally, our results present an interesting counterpoint to recent proofs of eigenvector-eigenvalue identity proven for Hermitian matrices \cite{Denton2019} and suggest that information parity between eigenvalues and eigenvectors may not hold for open system physics described by complex symmetric matrices, even in principle. 
%
\begin{acknowledgments}
The authors wish to thank John Teufel, Hakan E. T\"{u}reci and Emery Doucet for useful conversations, and Tristan Brown for comments on the manuscript. This research was supported by the U.S. Department of Energy under grant numbers DE-SC0019515 (C.K.) and DE-SC0019461 (Z.X.). AM acknowledges funding by the Deutsche Forschungsgemeinschaft through the Emmy Noether program (Grant No. ME 4863/1-1) and the project CRC 910.
\end{acknowledgments}
%
\appendix
%
\section{Eigenvalue analysis for three- and four-mode systems}
\label{App:Evalanalyis}
%
\subsection{$N=3$ case}
%
We first write the characteristic polynomial $p(\lambda)$ of $\mathbb{M}^{(3)}$, as $p(\lambda) = \alpha \lambda^{3} + \beta \lambda^{2} + \gamma \lambda + \delta$, where
\begin{subequations}
\begin{align}
    \alpha &= 1, \\
    \beta &= \frac{\kappa_1 + \kappa_2 + \kappa_3}{2}, \\
    \gamma &= g_1^{(3)2} + g_2^{(3)2} + \frac{\kappa_1\kappa_2 + \kappa_1\kappa_3 + \kappa_2\kappa_3}{4}, \\
    \delta &= \frac{4g_1^{(3)2}\kappa_3 + 4g_2^{(3)2}\kappa_1 + \kappa_1\kappa_2\kappa_3}{8}.
\end{align}
\end{subequations}
Using $\textit{Cardano}$'s method, we can first write $p(x)$ into the depressed cubic form $p^{\prime}(t)$ by substituting $\lambda = t - \beta/3\alpha$, such that
\begin{align}
    p^{\prime}(t) &= t^{3} + 3\epsilon_{1}t + 2\epsilon_{2},
\end{align}
with
\begin{subequations}
\begin{align}
    & \epsilon_{1} = \frac{3\alpha\gamma - \beta^{2}}{9\alpha^{2}}; \\
    & \epsilon_{2} = -\frac{9\alpha\beta\gamma - 27\alpha^{2}\delta - 2\beta^{3}}{54\alpha^{3}}.
\end{align}
    \label{Eq:cubictodepcubic}
\end{subequations}
Solving for the roots of the cubic equation, $p^{\prime}(t) =0$, and using Eqs.~(\ref{Eq:cubictodepcubic}), gives the eigenvalues $e_i^{(3)}$ of $\mathbb{M}^{(3)}$ as
\begin{subequations}
\begin{align}
e_1^{(3)} &= \eta_{0} + \eta_{+} + \eta_{-} , \\
e_2^{(3)} &= \eta_{0} + e^{i2\pi/3}\eta_{+} + e^{i4\pi/3}\eta_{-}, \\
e_3^{(3)} &= \eta_{0} + e^{i4\pi/3}\eta_{+} + e^{i2\pi/3}\eta_{+},
\end{align}
\end{subequations}
where $\eta_{0} = -\beta/(3\alpha), \; \eta_{\pm} = \left(\epsilon_{1}\pm \sqrt{\epsilon_{1}^{2} + \epsilon_{2}^{3}}\right)^{1/3}$. In this representation, the location of exceptional points can be found as \cite{AmShallem2015}
\begin{eqnarray}
    & & \text{EP2}: \text{disc}(\mathbb{M}^{\rm (3)}) = 0 \iff \epsilon_{1}^{2} +\epsilon_{2}^{3} = 0\\
    & & \text{EP3}: \epsilon_{1} = 0 \; \text{and} \; \epsilon_{2} = 0.
\end{eqnarray}
%
\subsection{$N=4$ case}
%
For the 4-mode case, we similarly write the characteristic polynomial of $\mathbb{M}^{(4)}$ as $p(\lambda) = a\lambda^{4} + b\lambda^{3} + c\lambda^{2} + d\lambda + e$, where 
%
\begin{subequations}
\begin{align}
a &=\,1, \\
b &=\,\frac{\kappa_1 + \kappa_2 + \kappa_3 + \kappa_4}{4}, \\
c &=\,2g_1^{(4)2} + g_2^{(4)2} \nonumber\\ 
& \quad +\frac{\kappa_1\kappa_2 + \kappa_1\kappa_3+\kappa_2\kappa_3+\kappa_1\kappa_4+\kappa_2\kappa_4+\kappa_3\kappa_4}{4},\\
d &=\,\frac{1}{8}\Big(4g_1^{(4)2}(\kappa_1 + \kappa_2 + \kappa_3 + \kappa_4) + 4g_2^{(4)2}(\kappa_1 + \kappa_4) \nonumber\\
& \quad \quad  \quad + \kappa_1\kappa_2\kappa_3 + \kappa_1\kappa_2\kappa_4 + \kappa_1\kappa_3\kappa_4 + \kappa_2\kappa_3\kappa_4\Big),\\
e &=\,g_1^{(4)4} + \frac{1}{4}g_1^{(4)2}\left(\kappa_1\kappa_2 + \kappa_3\kappa_4\right) \nonumber\\
& \qquad \qquad + \frac{1}{16}\kappa_1\kappa_4\left(4g_2^{(4)2} + \kappa_2\kappa_3\right). 
\end{align}
\end{subequations}
Using $\textit{Ferrari}$'s method, we rewrite $p(\lambda)$ in depressed quartic form $P^{\prime}$ by substituting $\lambda = y - b/(4a)$ such that
\begin{align}
    P^{\prime}(y) &= y^{4} + f_{1} y^{2} + f_{2} y + f_{3},
\end{align}
where
\begin{subequations}
\begin{align}
    f_{1}  &= \frac{8ac - 3b^{2}}{8a^{2}}, \\
    f_{2} &= \frac{b^{3} - 4abc+ 8a^{2}d}{8a^{3}}, \\
    f_{3} &= \frac{-3b^{4}+16ab^{2}c-64a^{2}bd+256a^{3}e}{256a^{4}}.
\end{align}
\end{subequations}
Solving for $y$, and subsequently $\lambda$, gives the eigenvalues of $\mathbb{M}^{(4)}$ as
\begin{subequations}
\begin{align}
    e_1^{(4)} &= G_{1} - G_{3} + \sqrt{-G_{3}^{2}-\frac{f_{1}}{2}+\frac{f_{2}}{4G_{3}}}, \\
    e_2^{(4)} &= G_{1} - G_{3} - \sqrt{-G_{3}^{2}-\frac{f_{1}}{2}+\frac{f_{2}}{4G_{3}}}, \\
    e_3^{(4)} &= G_{1} + G_{3} + \sqrt{-G_{3}^{2}-\frac{f_{1}}{2}-\frac{f_{2}}{4G_{3}}},\\
    e_4^{(4)} &= G_{1} + G_{3} - \sqrt{-G_{3}^{2}-\frac{f_{1}}{2}-\frac{f_{2}}{4G_{3}}},
\end{align}
\end{subequations}
where
\begin{subequations}
\begin{align}
    G_{1} &= -\frac{b}{4a},\\
    G_{2} &= \left(\frac{g_1 + \sqrt{g_1^{2} - 4g_2^{3}}}{2}\right)^{1/3},\\
    G_{3} &= \frac{1}{2}\sqrt{g_{3}+ \frac{1}{3}\left(G_{2} + \frac{g_2}{G_{2}}\right)} 
\end{align}
    \label{Eq:J}
\end{subequations}
with
\begin{subequations}
\begin{align}
    g_1 &= \frac{2c^{3} - 9bcd + 27b^{2}e + 27ad^{2} - 72ace}{a^{3}}, \\
    g_2 &= \frac{c^{2} - 3bd + 12ae}{a^{2}},\\
    g_{3} & =\frac{3b^{2} - 8ac}{12 a^{2}} .
\end{align}
\end{subequations}
\begin{figure}[t!]
\centering
\includegraphics[width=0.9\columnwidth]{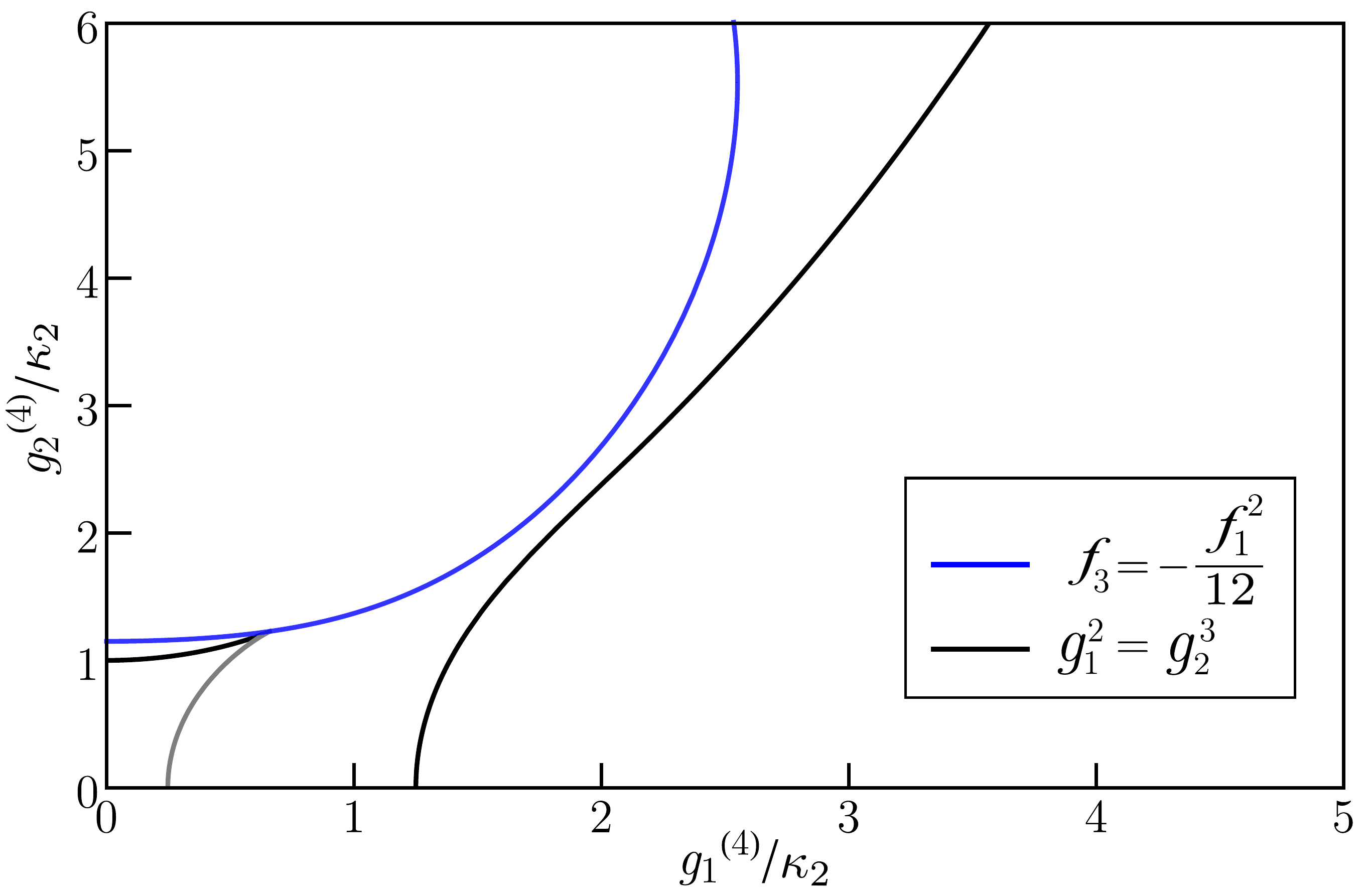}
\caption{Black curves denote locus of EP2 points obtained using ${\rm disc}(\mathbb{M}^{(4)}) = 0$, while blue curve is a parametric plot of the second condition in Eq. (\ref{Eq:EP3in4mode}) calculated for the decay rates used in Fig.~\ref{Fig:OpenLoop}(c). The intersection of the black and blue curves gives the location of EP3 for the four-mode system.}
\label{Fig:EP3in4mode}
\end{figure}
In this representation, the location for exceptional points follows similarly as in the $N=3$ case
\begin{eqnarray}
    & & \text{EP2}: \text{disc}(\mathbb{M}^{\rm (4)}) = 0 \iff g_{1}^{2} - 4g_{2}^{3} = 0.
\end{eqnarray}
However, for EP3 in the 4-mode system,
\begin{eqnarray}
g_{1}^{2} = 4g_{2}^{3} \quad {\rm and} \quad f_{3} = -f_{1}^{2}/12.
\label{Eq:EP3in4mode}
\end{eqnarray}
We note that for the decay rates used in the main text, exactly one EP3 is realized for the ${N=4}$ system as depicted in Fig.~\ref{Fig:EP3in4mode}.
%
%
\section{Calculations for three-mode cooling}
\label{App:full3mode}
%
Using the Hamiltonian in Eq.~(\ref{Eq:Hamiltonian}) for $N=3$, we can write the equations of motion for the three-mode optomechanical system as
\begin{align}
\frac{d\hat{a}_{1}}{dt} &= -ig_{1}\hat{a}_{2}-\frac{\kappa_{1}}{2}\hat{a}_{1}+\sqrt{\kappa_{1}}\hat{a}_{1}^{\text{in}}, \\
\frac{d\hat{a}_{2}}{dt} &= -ig_{1}\hat{a}_{1}-ig_{2}\hat{a}_{3}-\frac{\kappa_{2}}{2}\hat{a}_{2}+\sqrt{\kappa_{2}}\hat{a}_{2}^{\text{in}}, \\
\frac{d\hat{a}_{3}}{dt} &= -ig_{2}\hat{a}_{2}-\frac{{\kappa}_{3}}{2}\hat{a}_{3}+\sqrt{\kappa_{3}}\hat{a}_{3}^{\text{in}}.
\end{align}
This system of coupled differential equations can be solved as a system of algebraic equations in Fourier domain to obtain the solution for the mechanical mode operator $a_{1}[\omega]$,
\begin{align}
\hat{a}_{1}[\omega] &= \sqrt{\kappa_{1}}\left\{\frac{\left(g_{2}^{2}+\chi_{2}^{-1}\chi_{2}^{-1,{\rm eff}}\right)}
{g_{1}^{2}\chi_{2}^{-1}+\chi_{1}^{-1}
\left(g_{2}^{2}+\chi_{2}^{-1}\chi_{2}^{-1,{\rm eff}}\right)}\right\} \hat{a}_{1}^{\text{in}}[\omega] \nonumber \\
&+ \sqrt{\kappa_{2}}\left\{\frac{-ig_{1}\chi_{2}^{-1}}
{g_{1}^{2}\chi_{2}^{-1}+\chi_{1}^{-1}
\left(g_{2}^{2}+\chi_{2}^{-1}\chi_{2}^{-1,{\rm eff}}\right)}\right\}
\hat{a}_{2}^{\text{in}}[\omega] \nonumber\\
&+ \sqrt{\kappa_{3}}\left\{\frac{-g_{1}g_{2}}
{g_{1}^{2}\chi_{2}^{-1}+\chi_{1}^{-1}
\left(g_{2}^{2}+\chi_{2}^{-1}\chi_{2}^{-1,{\rm eff}}\right)}\right\}
\hat{a}_{3}^{\text{in}}[\omega], \nonumber\\
\end{align}
with the susceptibilities $\chi_{2}^{-1} = -i\omega + \kappa_{2}/2$, ${\chi_{2}^{-1,{\rm eff}} = -i\omega + \kappa_{2}^{\rm eff}/2}$, where $\kappa_{2}^{\rm eff} = \kappa_{2}(1+\mathcal{C}_{2})$. Using this we can calculate the symmetrized spectral density of the mechanical mode, 
\begin{eqnarray}
	\bar{S}_{\hat{a}_{1}\hat{a}_{1}}^{(3)}[\omega] & = & \frac{1}{2} \int_{-\infty}^{\infty} d\omega' \left\langle\hat{a}_{1}^{\dagger} [\omega] \hat{a}_{1}[\omega'] + \hat{a}_{1}[\omega] \hat{a}_{1}^{\dagger} [\omega']\right\rangle, \nonumber
\end{eqnarray}
from which effective population of the mechanical mode then follows as
\begin{eqnarray}
	\left(n_{1}^{(3)} + \frac{1}{2} \right) = \int\frac{d\omega}{2\pi}\bar{S}_{\hat{a}_{1}\hat{a}_{1}} [\omega].\nonumber
\end{eqnarray}
The resultant expression of $\bar{n}_{1}^{(3)}$ obtained following this procedure is
\begin{align}
   n_{1}^{(3)} &= n_{m}\frac{\sigma_1}{\Sigma} + n_{o}\frac{\sigma_2}{\Sigma} + n_{a}\frac{\sigma_3}{\Sigma}, 
   \label{Eq:3modefullexp}
\end{align}
where
\begin{subequations}
\begin{align}
\sigma_{1} &= 
\left[\frac{\kappa_{1}^{2}}{\kappa_{\parallel}^{2}} + \frac{\kappa_{1}^{3}}{\kappa_{2}\kappa_{\parallel}\kappa_{\perp}} + \frac{\kappa_{1}^{3}}{\kappa_{3}\kappa_{\parallel}\kappa_{\perp}} (1+\mathcal{C}_{1})\right](1+\mathcal{C}_{1} + \mathcal{C}_{2}) \nonumber\\
& \qquad + \left(\frac{\kappa_{1}}{\kappa_{\parallel}}\right)(1+\mathcal{C}_{2})^{2} 
- \left(\frac{\kappa_{1}^{2}}{\kappa_{\parallel}\kappa_{\perp}}\right) \mathcal{C}_{1}(1+\mathcal{C}_{2}),  \\
\sigma_{2} &= \left(\frac{\kappa_{1}}{\kappa_{\parallel}}\right)\mathcal{C}_{1} + \left(\frac{\kappa_{1}^{2}}{\kappa_{3}^{2}} \right)\mathcal{C}_{1}(\mathcal{C}_{1}+\mathcal{C}_{2}),\\
\sigma_{3} &= \left(\frac{\kappa_{1}}{\kappa_{\parallel}} + \frac{\kappa_{1}^{2}}{\kappa_{\parallel}\kappa_{\perp}}\right) \mathcal{C}_{1}\mathcal{C}_{2},\\
\Sigma &= (1+\mathcal{C}_{1} + \mathcal{C}_{2})\left[\frac{\kappa_{1}}{\kappa_{\parallel}}(1+\mathcal{C}_{2}) + \frac{\kappa_{1}^{2}}{\kappa_{3}^{2}}(1+\mathcal{C}_{1}) \right],
\end{align}
\end{subequations}
with $\kappa_{\parallel} = \kappa_{2}\kappa_{3}/(\kappa_{2}+\kappa_{3}), \; \kappa_{\perp} =\kappa_{2}+\kappa_{3}$.
%
%
%

%
\end{document}